\documentclass[twocolumn,showpacs,preprintnumbers,amsmath,aps,amssymb]{revtex4}
\usepackage{amsmath}
\usepackage{amsfonts}
\usepackage{amssymb}
\usepackage{graphicx}
\providecommand{\U}[1]{\protect\rule{.1in}{.1in}}

\begin{document}
\title{Towards Noncommutative Quantum Black Holes}
\author{J. C. L\'opez-Dom\'{\i}nguez}
\email{jlopez@fisica.ugto.mx}
\author{O. Obreg\'on}
\email{octavio@fisica.ugto.mx}
\author{M. Sabido}
\email{msabido@fisica.ugto.mx}
\affiliation{Instituto de F\'{\i}sica de la Universidad de Guanajuato P.O. Box E-143, 37150
Le\'on Gto., M\'exico}
\author{C. Ram\'{\i}rez}
\email{cramirez@fcfm.buap.mx}
\affiliation{Facultad de Ciencias F\'{\i}sico Matem\'aticas, Universidad Aut\'onoma de
Puebla, P.O. Box 1364, 72000 Puebla, M\'exico.}
\date{\today}

\begin{abstract}
In this paper we study noncommutative black holes.
We use a diffeomorphism between the
Schwarzschild black hole and the Kantowski-Sachs cosmological model,
which is generalized to noncommutative minisuperspace. Through the
use of the Feynman-Hibbs procedure we are able to study the
thermodynamics of the black hole, in particular, we calculate the
Hawking's temperature and entropy for the noncommutative
Schwarzschild black hole.

\end{abstract}
\maketitle

\section{Introduction}
The search for a quantum theory of gravity has been a long and difficult one; a direct
quantization of general relativity using the tools of quantum field theory gives a theory
with an ill ultraviolet behavior, and a lack of a fundamental physical principle to construct
the theory makes matters worst. One can think that  the black hole plays a similar role in
quantum gravity as the atom played in the development of quantum mechanics and
quantum field theory.  We may use
it as a starting point for testing different constructions that include quantum aspects of
gravity. Research on black hole physics has uncovered several mysteries: Why is the statistical
black hole entropy proportional to the horizon area? What happens to the information in black
hole evaporation? The answer to these questions and others have been extensively studied in the
literature particularly in the two main proposals to build a quantum theory of gravity, namely,
string theory \cite{horo}, and loop quantum gravity \cite{ash}.

Since the birth of general relativity people started searching for
solutions to Einstein's field equations and very powerful and
sophisticated methods have been developed. For some time it has been
known that changing the causal structure of space time (i.e.
interchanging the coordinates $r\leftrightarrow t$), changes a
static solution for a cosmological one. The best known case is the
Schwarzschild metric that under this particular diffeomorphism
transforms into the Kantowski-Sachs metric \cite{ro}. This interchange of
variables has been used recently as a method to generate new cosmological solutions
\cite{kergow,revqm,om1}. Later in an independent
way in \cite{qv1,qv2}, it has been suggested that
by interchanging $r \leftrightarrow it$ we can get time dependent solutions from stationary solutions
in string theory. In this way we may relate $Dp$-brane (black hole
like) solutions to $S$-brane solutions, i.e. time dependent backgrounds
of the theory. On the other side, there are proposals to
obtain $S$-brane solutions \cite{om1,Peet,qo,om2}; thus, if cosmological
solutions ($S$-branes) can be generated from
stationary ones ($Dp$-branes), we can expect that this procedure works the other way
around.

In the past a lot of work has been done in relation with cosmology and
quantum cosmology \cite{lmike,hh}, so one can expect that
the classical exchange of $t \leftrightarrow r$ can be applied at
quantum level, that is, use the WDW (Wheeler-DeWitt) equation of the
cosmological (time dependent) models and from it obtain the
corresponding quantum equation for the stationary ones.

In the last years, noncommutativity (NC) has attracted a lot of attention
\cite{connes,sw}. Although most of
the work has been in the context of Yang-Mills theories, noncommutative
deformations of gravity have been proposed \cite{mofat,cham}. In
\cite{sab1}, the
Seiberg-Witten map is used consistently to write a
noncommutative theory of gravity, which only depends on the
commutative fields and their derivatives. If we attempt
to write down the field equations and solve them, it turns to be technically very difficult, due to
the highly non linear character of the theory. In \cite{ncqc}, an
alternative to incorporate noncommutativity to cosmological models has been proposed, by
performing a noncommutative deformation of the minisuperspace. The
authors applied this idea to the Kantowski-Sachs metric, and were able
to find the exact wave function for the noncommutative model.

Further, we know that from quantum mechanics we can get the
thermodynamical properties of a system. This already has been used
in connection with black holes \cite{york}. In \cite{paths}
the authors use the Feynman-Hibbs path integral procedure \cite{feynman}, to calculate
the thermodynamical properties, temperature and entropy, of a black hole, in agreement
with previous results \cite{otros}.

In this paper we apply some of these
ideas to obtain thermodynamical properties for a quantum black hole
and its noncommutative counterpart. We propose a quantum
equation for the Schwarzschild black hole, starting from the WDW equation for
the Kantowski-Sachs cosmological model.
We  apply the Feynman-Hibbs method to calculate the thermodynamical
properties. As we will show the temperature of the black hole and
its entropy get corrected, one of the corrections to the entropy is
a logarithmic correction that already has been obtained by other
means \cite{otros}. We  extend this procedure to include
noncommutativity by making the same kind of ansatz as in
\cite{ncqc}, namely, imposing that the minisuperspace variables do
not commute; from this we  are able to define the WDW equation for
the noncommutative Schwarzschild black hole  and following a similar
procedure as in \cite{paths}, we find the noncommutative wave function, the
temperature and entropy of the {\it``noncommutative Schwarzschild
black hole"}. As we will see apart of the presence of the factor $e^{-3\nu\theta}$
that multiplies the commutative entropy and temperature, the functional form of
the corrections are of the same kind. We
also show how the minisuperspace coordinate $\Omega$ is
changed by noncommutativity, allowing us to infer the
noncommutative entropy, if one uses the well-known Euclidean
calculation. The result agrees with the most relevant term of the
entropy calculated from the noncommutative WDW equation and the
Feynman-Hibbs procedure followed here.

The paper is organized as
follows: In section II we exhibit the WDW equation for the
Schwarzschild metric and the corresponding wave function. In section
III we review the proposal of introducing noncommutativity in the
minisuperspace, and obtain the noncommutative Wheeler-DeWitt (NC-WDW) equation
for the Schwarzschild black hole. In section IV we use the Feynman-Hibbs method on the WDW equation to
calculate the entropy of the Schwarzschild black hole, and apply the method to calculate the entropy of the noncommutative black hole. And finally section V
is devoted to conclusions and outlook.


\section{A Wheeler-DeWitt equation for the Schwarzschild Black Hole}

Let us begin by reviewing the relationship between the cosmological
Kantowski-Sachs metric and the Schwarzschild metric \cite{ro}. The
Schwarzschild solution can be written as
\begin{eqnarray}
ds^{2}=&-&\left(1-\frac{2m}{r}\right)dt^{2}+\left( 1-\frac{2m}%
{r}\right)^{-1}dr^{2}\nonumber\\
&+&r^{2}\left(d\theta^{2}+\sin^{2}\theta d\varphi^{2}\right) .
\end{eqnarray}
For the case $r<2m$, the $g_{tt}$ and $g_{rr}$ components of the metric change
in sign and $\partial_{t}$ becomes a spacelike vector. If we make the
coordinate transformation $t\leftrightarrow r$, we find%
\begin{eqnarray}
ds^{2}=&-&\left(\frac{2m}{t}-1\right)^{-1}  dt^{2}+\left(  \frac{2m}%
{t}-1\right)dr^{2}\nonumber \\
&+&t^{2}\left(d\theta^{2}+\sin^{2}\theta d\varphi^{2}\right),
\end{eqnarray}
when compared with the parametrization by Misner of the Kantowski-Sachs metric%
\begin{eqnarray}
ds^{2}=&-&N^{2}dt^{2}+e^{\left( 2\sqrt{3}\gamma\right)} dr^{2} \nonumber\\
&+&e^{\left(-2\sqrt{3}\gamma\right)} e^{\left(-2\sqrt{3}\Omega\right)}
\left(d\theta^{2}+\sin^{2}\theta d\varphi^{2}\right),
\label{KSmetric}%
\end{eqnarray}
we identify
\begin{gather}
N^{2}=\left(\frac{2M}{t}-1\right)^{-1},~~~~~~
e^{2\sqrt{3}\gamma}=\frac{2M}{t}-1,\nonumber \\
e^{-2\sqrt{3}\gamma}e^{-2\sqrt{3}\Omega}=t^{2},
\label{dif}
\end{gather}
where this metric with the identification of the $N$, $\gamma$, and $\Omega$
functions  is also a classical solution for the Einstein
equations.
The metric (\ref{KSmetric}) can be introduced into de ADM (Arnowitt-Deser-Misner)
action and a consistent set of equations for $N$, $\gamma$, and $\Omega$ can be
obtained by varying the action, these equations can be show to be
equivalent to the Einstein equations. The
corresponding Wheeler-DeWitt equation for the
Kantowski-Sachs metric, with some particular factor ordering, is%
\begin{equation}
\left[  -\frac{\partial^{2}}{\partial\Omega^{2}}+\frac{\partial^{2}}%
{\partial\gamma^{2}}+48e^{ -2\sqrt{3}\Omega } \right]
\psi(\Omega,\gamma)=0.\label{ks}
\end{equation}
The solution of this equation is given by \cite{misner}
\begin{equation}
\psi_\nu=e^{\pm i\nu\sqrt{3}\gamma}K_{i\nu}\left(4e^{-\sqrt{3}\Omega}\right) ,
\end{equation}
where $\nu$ is the separation constant and $K_{iv}$ are the modified
Bessel functions.
In \cite{ro1} it has been shown that this wave function describes quantum planck size states.


\section{NonCommutative Quantum Cosmology and the Quantum Black Hole}

The noncommutative deformation for this cosmological model, has been proposed
in \cite{ncqc}.
We begin by modifying the simplectic structure in minisuperspace, by assuming that the
coordinates $\Omega$ and $\gamma$ obey the commutation relation%
\begin{equation}
\left[  \Omega,\gamma\right] = i\theta,
\label{non}
\end{equation}
in a similar fashion as in noncommutative quantum mechanics. As usual this deformation
can be reformulated in terms of the Moyal product%
\begin{eqnarray}
&f&\left( \Omega,\gamma\right) \ast g\left( \Omega,\gamma\right)\nonumber \\
&=&f\left( \Omega,\gamma\right) e^{\left(  i\frac{\theta}{2}\left(
\overleftarrow
{\partial}_{\Omega}\overrightarrow{\partial}_{\gamma}-\overleftarrow{\partial
}_{\gamma}\overrightarrow{\partial}_{\Omega}\right)\right)} g\left(
\Omega,\gamma\right)  .
\end{eqnarray}
Now the Wheleer-DeWitt equation for the noncommutative theory will be%
\begin{equation}
\left[-P_{\Omega}^{2}+P_{\gamma}^{2}-48e^{-2\sqrt{3}\Omega}\right]
\ast\psi(\Omega,\gamma)=0.
\end{equation}
As is know in noncommutative quantum mechanics, the original
phase space is modified. It is possible to reformulate in terms
of the commutative variables and the ordinary product of functions, if
the new variables $\Omega\rightarrow\Omega+\frac{1}{2}\theta
P_{\gamma}$ and $\gamma \rightarrow\gamma-\frac{1}{2}\theta P_{\Omega}$ are
introduced. The momenta remain the same. As a consequence, the
original WDW equation changes, with a modified
potential  $V(\Omega,\gamma)$\cite{mezincescu},
\begin{eqnarray}
V\left( \Omega,\gamma\right) \ast\psi\left( \Omega,\gamma\right) &=&V\left(
\Omega-\frac{\theta}{2} P_{\gamma},\gamma+\frac{\theta}{2}P_{\Omega}\right)\nonumber\\
&\times&\psi\left(\Omega,\gamma\right),
\end{eqnarray}
so the NC-WDW equation takes the form%
\begin{equation}
\left[  -\frac{\partial^{2}}{\partial\Omega^{2}}+\frac{\partial^{2}}%
{\partial\gamma^{2}}+48e^{\left( -2\sqrt{3}\Omega+\sqrt{3}\theta
P_{\gamma }\right) } \right]  \psi(\Omega,\gamma)=0. \label{ncwdw}
\end{equation}
We solve this equation by separation of variables with the ansatz%
\begin{equation}
\psi(\Omega,\gamma)=e^{i\sqrt{3}\nu\gamma } \chi\left(
\Omega\right),
\end{equation}
where $\sqrt 3 \nu$ is the eigenvalue of $P_{\gamma}$.
Thus $\chi\left(  \Omega\right)  $ satisfies the equation \cite{ncqc},%
\begin{equation}
\left[ -\frac{d^{2}}{d\Omega^{2}}+48e^{\left(
-2\sqrt{3}\Omega+3\nu\theta\right)}-3\nu^2\right]\chi(\Omega)=0.\label{ncbh}
\end{equation}
The solution of the NC-WDW equation is%
\begin{equation}
\psi_{\nu}(\Omega,\gamma)=e^{\left(i\sqrt{3}\nu\gamma\right)}
K_{i\nu}\left[  4e^{-\sqrt{3}\left( \Omega- \sqrt{3}
\nu\theta/2 \right) } \right]  .
\end{equation}
In \cite{ncqc} the consequences of this wave function have been analyzed.
As a result of noncommutativity the
probability density has several maxima, which correspond to
new stable states of the Universe, opposite to the commutative
case where only one stable state exists. Following the previous section,
this wave function could describe noncommutative quantum black holes.
We can find the noncommutative temporal evolution of $\Omega$ and $\gamma$
trough a WKB type method; this yields the classical noncommutative
solutions of the Kantowski-Sachs universe \cite{mena,bb}.

\section{NonCommutative Black Hole Entropy}
In this section we compute the temperature and entropy for the
black hole using the Feynman-Hibbs procedure for
statistical mechanics, and then we apply it to the noncommutative
black hole.

Following Section II, we consider Eq. (\ref{ks}). This equation depends on
two variables $(\Omega,\gamma)$, but after the
separation of variables we have $\psi(\Omega,\gamma)=e^{
i\sqrt{3}\nu\gamma}\chi(\Omega)$. Thus, the dependence on the variable $\gamma$
is the one of a plane wave and is eliminated when computing the thermodynamical
observables. Therefore, we could consider this as a suitable approach for the black hole,
described by the following equation:
\begin{equation}
\left[-\frac{d^{2}}{d\Omega^{2}}+48 e^{-2\sqrt{3}\Omega}\right]\chi(\Omega)
=3\nu^2 \chi(\Omega).\label{bh}
\end{equation}
Now that we have this quantum equation, we can use the Feynman-Hibbs procedure
to compute the partition function of the black hole. This procedure has the advantage that
the relevant information is contained in the potential function
\begin{equation}
V(\Omega)=48 e^{-2\sqrt{3}\Omega}.
\end{equation}
This exponential potential can always be expanded, in particular, the calculations are simplified for small $\Omega$. After  expanding to second
order in $\Omega$,  we make the change of variable $\alpha=\sqrt{6}\Omega-1/\sqrt{2}$,
multiply Eq. (\ref{bh}) by $\frac{E_p}{12}$, and finally rename
$\alpha=\frac{x}{l_p}$. We get the familiar form,
\begin{equation}
\left[-\frac{1}{2}l_{p}^{2}E_{p}\frac{d^{2}}{dx^{2}}+4\frac{E_{p}}{l_{p}%
^{2}} x^{2}\right] \chi\left( x\right) = E_{p}\left[
\frac{\nu^2}{4}-2\right] \chi\left( x\right). \label{osc}
\end{equation}
When comparing with the usual harmonic oscillator we identify
$\hbar\omega=\sqrt{\frac{3}{2\pi}}E_{p}$  in order to
obtain the correct Hawking temperature for the black hole \footnote{
The factor corresponds to an ambiguity in the value of the lowest eigenvalue of the spectrum
\cite{kastrup}. In loop quantum gravity the Barbero-Immirizi parameter \cite{barbero} is fixed
in a similar way.}.

Further, the Feynman-Hibbs procedure allows to incorporate the quantum
corrections to the partition function through the {\it
``corrected"} potential, which results in \cite{feynman}
\begin{equation}
U(x)=\frac{3E_{p}}{4\pi l_{p}^{2}}
\left[x^{2}+\frac{\beta l_{p}^2E_{p}}{12}\right].
\end{equation}
Thus, the corrected partition function is given by
\begin{equation}
Z_{Q}=\sqrt{\frac{3}{2\pi}}\frac{e^{ -\frac{\beta^{2}%
E_{p}^{2}}{16\pi}}}{\beta E_{p}},
\label{part}
\end{equation}
from which we can proceed to calculate the thermodynamics of interest.
The internal energy of the black hole is
\begin{eqnarray}
\bar{E} &=&-\frac{\partial}
{\partial\beta}\ln Z_{Q}\nonumber \\
& = &\frac{1}{8\pi}\beta E_{p}^{2}+\frac {1}{\beta}=Mc^{2}.
\end{eqnarray}
Solving for $\beta$ in terms of the
Hawking temperature $\beta_{H}=\frac{8\pi Mc^{2}}{E_{p}^{2}}$, the
corrected temperature of the black hole is
\begin{align}
\beta &= \frac{8\pi Mc^{2}}{E_{p}^{2}}\left[
1-\frac{1}{8\pi}\left(\frac{E_{p}}{Mc^{2}}\right)^{2}\right]
=\beta_{H} \left[1-\frac{1}{\beta_{H}} \frac{1}{Mc^{2}}\right].
\label{24}
\end{align}
In order to calculate the entropy we use
\begin{equation}
\frac{\mathit{S}}{k}=\ln{Z_Q}+\beta\bar{E},
\end{equation}
from which
we arrive to the corrected entropy. In terms of the Bekenstein-Hawking entropy
$\frac{\mathit{S_{BH}}}{k}=4\pi\left(\frac{Mc^2}{E_p}\right)^2=\frac{A_s}{4l_{p}^2}$,
the black hole entropy takes the simple form,
\begin{equation}
\frac{\mathit{S}}{k}=\frac{\mathit{S_{BH}}}{k}-\frac{1}{2}\ln{\left[
\frac{\mathit{S_{BH}}}{k}\right]}+\mathcal{O}\left({\mathit{S}^{-1}_{BH}}\right).
\end{equation}
This result has the interesting feature that the coefficient of the first
correction, the logarithmic one, agrees with the one obtained in string theory
\cite{ads}, as well as in loop quantum gravity \cite{loop1}. The form
of this correction had been known already from other works \cite{otros},
where it was obtained by other considerations.

In order to calculate the thermodynamics of the noncommutative
black hole, we follow the same method as before, and we
start by considering Eq. (\ref{ncbh}). A straightforward calculation,
using the same steps as in the commutative
case, shows that Eq. (\ref{ncbh}) can be cast to the simple expression,
\begin{eqnarray}
&&\left( -\frac{1}{2} l_{p}^{2}E_{p}\frac{d^{2}}{dx^{2}}+4\frac{E_{p}}{l_{p}%
^{2}}e^{ 3\nu\theta}  x^{2}\right)
\chi\left(  x\right)\nonumber\\
&&=E_{p}\left(\frac{\nu^2}{4}-2e^{3\nu\theta} \right)
\chi\left( x \right).\label{ncosc}
\end{eqnarray}
Thus noncommutativity gives a modified version of Eq. (\ref{osc}),
with modified potential
\begin{equation}
V_{NC}(x)=4\frac{E_{p}}{l_{p}^{2}}e^{ 3\nu\theta} x^{2},
\end{equation}
and a ``frequency",
\begin{equation}
\hbar\omega_{NC}=\sqrt{\frac{3}{2\pi}} E_{p}e^{\frac{3\nu\theta}{2}
},\label{hwNC}
\end{equation}
which coincides with the commutative case for $\theta=0$.
Now we can directly apply the Feynman-Hibbs method to Eq.
(\ref{ncosc}) and obtain the corrected partition function,
\begin{equation}
Z_{NC}=\sqrt{\frac{2\pi}{3}}\frac{e^{-\frac{3\nu\theta}{2}}}{\beta E_{p}}{\exp{\left( -\frac{\beta^{2}
E_{p}^{2}e^{3\nu\theta}}{16\pi}\right)}} ,
\end{equation}
from which we calculate the temperature
of the noncommutative black hole in terms of the commutative Hawking
temperature,
\begin{equation}
\beta= \beta_{H}e^{-3\nu\theta}\left[ 1- \frac{1}{\beta_{H}}
\frac{e^{3\nu\theta}}{Mc^{2}} \right]\label{betaNC}.
\end{equation}
If we define the noncommutative Hawking
temperature $\beta_{H}^{NC}=\beta_{H}e^{-3\nu\theta}$, the black hole
temperature takes the same form as in the commutative case,
\begin{equation}
\beta=\beta_{H}^{NC}
\left[1-\frac{1}{\beta_{H}^{NC}}\frac{1}{Mc^2}\right].
\label{nct}
\end{equation}
The entropy is calculated following the previous steps.
If we define the noncommutative Hawking-Bekenstein entropy as
$S_{NCBH}=S_{BH}e^{-3\nu\theta}$, we get
\begin{equation}
\frac{\mathit{S_{NC}}}{k}=%
\frac{\mathit{S}_{NCBH}}{k}-\frac{1}{2}\ln\left[%
\frac{\mathit{S}_{NCBH}}{k}\right]+\mathcal{O}\left({\mathit{S}^{-1}_{NCBH}}\right).
\label{nce}
\end{equation}
It has the same form as the
commutative case; again the logarithmic correction to the entropy appears
with a $-\frac{1}{2}$ factor. Also it is clear that we get
the commutative entropy in the limit $\theta\to 0$.

We may infer some properties for the temperature and entropy, using
Eq(\ref{dif}), we arrive to
\begin{equation}
\left( 1-e^{-2\sqrt{3}\gamma}\right)e^{-\sqrt{3}\gamma}e^{-\sqrt{3}\Omega}=2M.\label{masa}
\end{equation}
We can construct a similar expression for the
noncommutative metric by replacing the functions
$\gamma$, $\Omega$, and the mass parameter $M$ by their
noncommutative counterparts \cite{bb} $\gamma_{NC}$, $\Omega_{NC}$, and
$M_{NC}$. To relate the noncommutative theory with
the commutative one, we substitute the classical solutions from \cite{mena,bb} in Eq. (\ref{masa})
and in its noncommutative version.
From this we can find a simple relationship between the masses
$M_{NC}=Me^{\sqrt{3}\theta P_{\gamma_0}}$. From the relation between
the temperature and entropy and the mass of the black hole \cite{hw}, we arrive to
\begin{eqnarray}
\beta_{NCBH}&=&\beta_{BH} e^{- \sqrt{3}\theta P_{\gamma_0}},\nonumber\\
\mathit{S}_{NCBH}&=&\mathit{S}_{BH}e^{-\sqrt{3}\theta P_{\gamma_0}}.
\end{eqnarray}
Thus, we have the same behavior for the noncommutative temperature and
entropy as in (\ref{nct}) and (\ref{nce}).

\section{Conclusions and Outlook}
Black holes are natural candidates to probe aspects of quantum gravity,
in particular, noncommutative effects.

In principle, in order to study noncommutative black holes,
we would require a noncommutative version of general relativity, and then solve its
field equations to find the noncommutative metric. This is a very
difficult task.

However, it is possible to
circumvent these difficulties as we show in this paper.
With this idea in mind, we have extended the proposal
of noncommutativity in \cite{ncqc} to the Schwarzschild black hole
by using the diffeomorphism between the Kantowski-Sachs and
Schwarzschild metrics. The corresponding NC-WDW equation
is used as the quantum equation for the noncommutative black hole.

We use the Feynman-Hibbs formalism to calculate the
thermodynamics of the Schwarzschild black hole and obtain the Hawking Bekenstein
temperature and then the entropy which has
the logarithmic correction with the coefficient $-1/2$. This value
has been recently predicted by string theory \cite{ads} and loop quantum gravity
\cite{loop1}.

Finally, these ideas are applied to the
noncommutative black hole, and the corresponding entropy and
temperature are obtained. As a result, these thermodynamic quantities
are modified due to the presence of the noncommutative parameter.
In particular, noncommutativity decreases the value of the entropy,
which can be understood from the fact that noncommutativity decreases the
available physical states.

These ideas could be applied to other singular static solutions, i.e. black hole
type solutions.

\acknowledgments{ \noindent This work was partially supported by grants
PROMEP-UGTO-CA-3, CONACYT 44515, CONACYT 51306 and CONACYT 47641.
M. S. is also supported by  PROMEP-PTC-085.}

\vskip 2truecm



\end{document}